# Strain engineering a persistent spin helix with infinite spin lifetime

Xue-Zeng Lu and James M. Rondinelli*

Department of Materials Science and Engineering, Northwestern University, Evanston, Illinois 60208, USA

**Email:** jrondinelli@northwestern.edu

**Abstract**

Persistent spin textures (PSTs) in solid-state materials arise from a unidirectional spin-orbit field in momentum space and offer a route to deliver long carrier spin lifetimes sought for future quantum microelectronic devices. Nonetheless, few three-dimensional materials are known to host PSTs owing to crystal symmetry and chemical requirements. There are even fewer examples demonstrated experimentally. Here we report that high-quality persistent spin textures can be obtained in the polar point groups containing an odd number of mirror operations. We use representation theory analysis and electronic structure calculations to formulate general discovery principles to identify PSTs hidden in known complex ternary layered and perovskite structures with large electric polarizations. We then show some of these materials exhibit PSTs without requiring any special crystalline symmetries. This finding removes the limitation imposed by mirror-symmetry protected PSTs that has limited compound discovery. Our general design approach enables the pursuit of persistent spin helices in materials exhibiting the $C_{3v}$ crystal class adopted by many quantum materials exhibiting large Rashba coefficients.

**Introduction**

A persistent spin texture (PST) in momentum space enables an infinite spin lifetime for a spin-wave mode called a persistent spin helix (PSH) [1–5], which makes it a useful spin texture to protect against spin-decoherence in spin field-effect transistors [6] and spin Hall effect applications [7]. In the PSH state, spin scattering is essentially quenched because the effective field governing the spin precession of the itinerant electrons is momentum independent. This feature leads to SU(2) symmetry of the spin components for the PSH, allowing precession in the same direction after a scattering event and potentially infinite spin lifetimes, because the SU(2) symmetry is robust to spin-independent disorder, including Coulomb and other many-body interactions [1].

In contrast, any effective momentum-dependent field removes SU(2) symmetry, such as Dresselhaus and Rashba interactions, permitting scattering effects to always reduce the spin lifetime through the Dyakonov-Perel (DP) spin relaxation mechanism at low temperatures [6]. Although the PST is a highly sought spin texture for the aforementioned reasons, few materials with PSTs in the bulk are known [8]. GaAs/AlGaAs [2,9] and InGaAs/InAlAs [10,11] heterostructures can exhibit PSTs by tuning the width of the quantum wells and carrier concentration; however, these artificial structures require atomic precision during growth and carefully controlled carrier densities. The PST mechanism in these main group semiconductor heterostructures arises from a balance between the strength of the Dresselhaus [12] and Rashba [13] spin splitting of the electronic bands, which are both correlated to the spin-orbital interaction (SOI) and require broken inversion symmetry [1,5,9–11], to impose an effective momentum-dependent magnetic field, $B(\mathbf{k}) \equiv \mathbf{B}(\mathbf{k}) \cdot \mathbf{S}$, on the spins.

Recently, another mechanism for PSTs was proposed [14–29], which although remains to be confirmed experimentally [30], permits bulk materials to exhibit the spin texture without requiring the balancing of Dresselhaus and Rashba interactions through interface design. This symmetry-protected PST (SP-PST) was predicted to occur when a nonsymmorphic symmetry operation (e.g., glide operation composed of a mirror plane and translation) commutes with the effective SOI field, $B(\boldsymbol{k})$, in a material without inversion symmetry [16]; however, as we demonstrated recently [31], this is a sufficient but not necessary condition. These minimal ingredients enable a *unidirectional* spin orientation at the band-edges in the electronic structure (spin-momentum locking), which in the absence of SOI would exist as a Kramers's degeneracy. We also demonstrated that any noncentrosymmetric material with high a symmetry $k$ point exhibiting $C_{2v}$ little group symmetry with an even number of mirror operations intersecting at that position in the Brillouin zone can be a potential PST material [31]. This understanding can guide identification of crystals exhibiting PSTs in other achiral polar groups. This finding should allow us to surpass the apparent limitation of PSTs appearing as recent serendipitous discoveries in monoclinic and orthorhombic crystal systems [20]. Indeed, high-quality PST materials in the monoclinic crystal system are still missing, and surprisingly, there are no reported PST materials from among the ferroelectric trigonal crystal system. Therefore, apart from the previous significant efforts in exploring PSTs in space groups with an even number of mirror symmetries, finding more materials/crystal classes with an odd number of mirror symmetries (i.e., monoclinic and trigonal crystal systems) showing PST would further enable experimental demonstration.

Here we use $\boldsymbol{k} \cdot \boldsymbol{p}$ models and first-principles calculations to show that PSTs can exist in polar structures of the trigonal crystal system. The PST is confined either along $k$ paths with (symmetry protection) or without (symmetry unconstrained) mirror elements. Our main finding from our $\boldsymbol{k} \cdot \boldsymbol{p}$ model is that PSTs occur at the conduction band edge, if the conduction band minimum appears near the midway point of the *k*-path across the high symmetry *k* point with $C_s$ symmetry (with 1 mirror operation, *m*) in the trigonal noncentrosymmetric crystal classes. We predict $Tl_3SbS_3$ and $LiNbO_3$ satisfy this requirement and exhibit PSTs. Furthermore, thin film versions of these two compounds subjected to coherent epitaxial constraints undergo a symmetry reduction from $C_{3v}$ crystal class with 3 mirror operations to $C_s$. The PSTs persist in thin films of these materials and their quality is enhanced, which we quantify using multiple criteria. The monoclinic *Cm* structure of $Tl_3SbS_3$ thin films supports a nearly perfect PST without any spin deviation that also simultaneously spans a large area of the Brillouin zone. These features enable access to the PST via through chemical doping. Our study provides a promising route to find the persistent spin helix in 3D polar phases with long spin lifetimes.

**Computational Methods**

Our total energy calculations were based on density functional theory (DFT) within the generalized gradient approximation (GGA) utilizing the revised Perdew-Becke-Erzenhof functional for solids (PBEsol) [32] implemented in the Vienna *Ab Initio* Simulation Package (VASP) [33–35]. We used a 550-eV plane wave cutoff energy for all calculations and the projector augmented wave (PAW) method [36] with Li 1*s* and 2*s* electrons, Rb 4*s*, 4*p* and 5*s* electrons, Tl 5*d*, 6*s* and 6*p* electrons, Nb 4*p*, 4*d* and 5*s* electrons, Ta 5*p*, 5*d* and 6*s* electrons, Sb 5*s* and 5*p* electrons, O 2*s* and 2*p* electrons, and S 3*s* and 3*p* electrons treated as valence states. Gaussian smearing (0.10 eV width) is used for the Brillouin-zone integrations. The *k*-point sampling was tested and converged for the different cells. The convergence thresholds for electronic relaxation and structure relaxation are $10^{-7}$ eV and -0.005 eV/Å, respectively. We tested the effects of SOC in the relaxation and found only tiny changes to the atomic structure; therefore, SOC effects were not included in structure relaxations.

**Results and discussion**

**Identification of PST in the trigonal system**

We consider trigonal systems, because they host common polar phases [23,37,38], such as in multiferroic BiFeO3 [39] and polar chalcogenides Ag3AsS3 [40] with $R3c$ symmetries. Furthermore, a recent report showed a rational way to discover materials with strong Rashba coefficients, where commonly identified materials with large Rashba coefficients were found to exhibit space groups in trigonal and hexagonal crystal systems and corresponding $C_{3v}$ or $C_{6v}$ point groups (i.e., 20 out of 34 space groups) [41]. This contrasts with 6 out of 34 space groups found in monoclinic and orthorhombic crystal systems having $C_{2v}$ or $C_s$ point groups. Therefore, it is important to formulate a strategy to search and sort which compounds in the trigonal crystal system will exhibit PSTs to enable their discovery.

The symmetry operations of the $R3c$ phase with $C_{3v}$ point group in Supplementary Table 1 allow us to deduce that there may be three unidirectional spin directions for the PST, which are along the [110], [100] and [010] directions of the hexagonal cell. The corresponding $k$-paths for the [110], [100] and [010] directions are $\{[k, -k, k_z], [1/2, 1/2, k_z], [-1/2, -1/2, k_z]\}$, $\{[k, -2k, k_z], [\mu, 0, k_z]\}$, and $\{[2k, -k, k_z], [0, \nu, k_z]\}$, respectively, where $\mu$ and $\nu$ can take values of ±1/2. If we further constrain our investigation to the $k_x - k_y$ plane, which will help keep the SU(2) symmetry and long spin lifetime under minimization of the commutator relations in Ref. [31], we find that the $k$-paths for the [110], [100] and [010] unidirectional spin directions are $[k, -k, 0]$, $[k, -2k, 0]$, and $[2k, -k, 0]$, respectively. Next, we assess whether symmetry requires the electronic bands to remain degenerate along these $k$ paths. For the $[k, -k, 0]$, $[k, -2k, 0]$, and $[2k, -k, 0]$ paths, the little-group symmetries of $\mathbf{k}$ are $m^r_{110}$, $m^r_{100}$ and $m^r_{010}$, respectively. Because there is no $\mathcal{TG}$ in the little group leading to $(\mathcal{TG})^2 \psi = -\psi$, there is no space-group symmetry protecting the Kramers degeneracy of the band edges by SOC effects for the three paths; therefore, the bands no longer touch along band trajectories in the $k_x - k_y$ plane.

Next, we derive $\mathbf{k} \cdot \mathbf{p}$ models for each $C_{3v}$ and $C_s$ symmetries, because a $k$-path with the mirror symmetry in the trigonal system, such as one of $[k, -k, 0]$, $[k, -2k, 0]$, and $[2k, -k, 0]$, will connect the $C_{3v}$ and $C_s$ $k$ points that we use to represent, for example, the zone center, Γ, and the zone boundary, Y, respectively. The detailed derivation of the $\mathbf{k} \cdot \mathbf{p}$ models are provided in the Supplementary Note 1 and Supplementary Tables 2-10. The $\mathbf{k} \cdot \mathbf{p}$ model with $C_{3v}$ symmetry, spin-orbital coupling terms up to the third order in $k$, and constrained in the $k_x - k_y$ plane is:

$$\mathcal{H}_\Gamma = \alpha_1 k_x \sigma_y + \alpha_2 k_y \sigma_x + \alpha_3 k_x \sigma_x + \alpha_4 k_y \sigma_y + \beta_1(k_x^3 - 3k_x k_y^2)\sigma_z + \beta_2(k_x^3 + k_x k_y^2)\sigma_y + \beta_3(k_y^3 + k_y k_x^2)\sigma_x + \beta_4(k_x^3 + k_x k_y^2)\sigma_x + \beta_5(k_y^3 + k_y k_x^2)\sigma_y \quad (1)$$

The same constraints applied to $C_s$ symmetry give:

$$\mathcal{H}_Y = \gamma_1 k_y \sigma_x + k_x(\gamma_2 \sigma_y + \gamma_3 \sigma_z) \quad (2)$$

where $\alpha$, $\beta$ and $\gamma$ are the spin-orbital coupling coefficients and using a Cartesian coordinate system. The matrix form for the wavefunctions including the spins for $\mathcal{H}_\Gamma$ and $\mathcal{H}_Y$ are:

$$\psi_\Gamma = \begin{pmatrix} \bar{E}^2 \\ \bar{E}^1 \\ \bar{E}_1^1 \\ \bar{E}_1^2 \end{pmatrix} \text{ and } \psi_Y = \begin{pmatrix} \bar{E}^2 \\ \bar{E}^1 \end{pmatrix} \quad (3)$$

where $\bar{E}^2$, $\bar{E}^1$ and $\bar{E}_1$ in $\psi_\Gamma$ are irreducible representations (*irreps*) of the double point group $3m$ (**Supplementary Note 1**). $\bar{E}_1^1$ and $\bar{E}_1^2$ are the two components of the two-dimensional *irrep* $\bar{E}_1$. $\bar{E}^2$ and $\bar{E}^1$ in $\psi_Y$ are *irreps* of the double point group $m$. Here, we only consider the symmetries of the wavefunctions and not their exact forms. $\psi_\Gamma$ transforms as a $D^{3/2}$ representation and $\begin{pmatrix}\bar{E}^2\\\bar{E}^1\end{pmatrix}$ transform as a $D^{1/2}$ representation. With the Clebsch–Gordan (C-G) coefficients, we can have the following for $\psi_\Gamma$:

$$\bar{E}^2 = \eta_1^1 A_1 \bar{E}^2 + \eta_2^1 E^1 \bar{E}_1^2 + \eta_3^1 E^2 \bar{E}_1^1$$

$$\bar{E}^1 = -\eta_1^1 A_1 \bar{E}^1 + \eta_2^1 E^1 \bar{E}_1^2 + \eta_3^1 E^2 \bar{E}_1^1$$

$$\bar{E}_1^1 = \eta_1^2 E^2 \bar{E}_1^2$$

$$\bar{E}_1^2 = \eta_2^2 E^1 \bar{E}_1^1 \qquad (4)$$

where $A_1$ and $E$ are the *irreps* of the double point group $3m$. $E^1$ and $E^2$ are the two components of the two-dimensional *irrep* $E$. Similarly, we obtain for $\psi_Y$:

$$\bar{E}^2 = \eta_1^3 A_1 \bar{E}^1 + \eta_2^3 A_2 \bar{E}_1^2$$

$$\bar{E}^1 = -\eta_1^3 A_1 \bar{E}^2 + \eta_3^3 A_2 \bar{E}_1^1 \qquad (5)$$

where $A_1$ and $A_2$ are the *irreps* of the double point group $m$ and $\eta$ is a constant.

The above results can be applied to any space group with $C_{3v}$ point symmetry. Through a detailed analysis of the polar structures of the trigonal system, we notice that there are two different trigonal polar space groups with the symmetries, that is:

Set 1: $P3m1$, $P31m$, $R3m$

Set 2: $P3c1$, $P31c$, $R3c$

In Set 1, there are no glide (translation plus mirror) operations in the primitive cell. In Set 2, there are nonsymmorphic (glide) symmetries. After we obtain the symmetries of the wavefunctions about the Γ point, it can be seen that in both wavefunctions represented by *irreps* $\bar{E}^2$ and $\bar{E}^1$, the $d$ orbitals can be occupied by $\frac{1}{2}$-spin up (i.e., $\bar{E}_1^2$) and $\frac{1}{2}$-spin down (i.e., $\bar{E}_1^1$) at the same time. Furthermore, there is no symmetry constraint on making the spin-dependent $d$ orbitals occupancies of $E^1$ (occupied by $\frac{1}{2}$-spin up) equal to $E^2$ (occupied by $\frac{1}{2}$-spin down). Therefore, if $k_y\sigma_y$ exists, such as along the $k_y$ path with mirror symmetry, then $\sigma_y$ cannot be cancelled by any of the $d$ orbitals of the same atom. Moreover, it is known that the spin in the $k_y$ path will be constrained to be along the $k_x$ direction by the mirror operation. So, $\sigma_y$ is either cancelled between two different atoms or there is no orbital having spin along $\sigma_y$. There are two atoms that can be connected by the mirror symmetry in $R3c$, therefore, if no $\sigma_y$ occurs along the $k_y$ path, there will remain orbitals having spins along $\sigma_y$ at each atom leading to a reduction of spin magnitude along $\sigma_x$ (i.e., the PST direction). In $R3m$, the mirror symmetry will transform one atom to itself, which will result in a situation in which there is no orbital having spin along $\sigma_y$. Thus, if there are PSTs in both $R3c$ and $R3m$, the quality of the PST as determined by the symmorphic mirror symmetry in $R3m$ should be better than that in $R3c$. Regarding the Y point, there is no $\sigma_y$ allowed in the $k_y$ path as indicated in Eq. (2), therefore, there is no significant reduction of the spin magnitude along $\sigma_x$ in $R3c$.

Next, we derive whether PSTs exists in the $R3c$ and $R3m$ space groups. From Eqs. (1) and (2), we can see the spin direction will be uniform, that is, along the $\sigma_x$ direction in the $k_y$ path, regardless of the high-symmetry point considered. This is the same result as that in $C_{2v}$, because of the mirror symmetry in the $k_y$ path. The main difference between $C_{3v}$ and $C_{2v}$ is that the spin deviation part led by $k_x$, which is present in the Hamiltonian having $C_{2v}$ symmetry and contributes to the term $k_x \sigma_y$ about both Γ and Y, adds additional terms of the form $\beta_1(k_x^3 - 3k_x k_y^2)\sigma_z + \beta_2(k_x^3 + k_x k_y^2)\sigma_y + \beta_5 k_y k_x^2 \sigma_y$ around Γ and $\gamma_3 k_x \sigma_z$ around Y in $C_{3v}$. These terms are responsible for the spin deviation. From Eqs. (1) and (2), the spin-deviation angles $(\theta_y, \theta_z)$ about Γ and Y with respect to $\sigma_x$ are

$$\left( \arctan\left(\frac{k_x(\alpha_1 + \beta_2 k_y^2)}{k_y(\alpha_2 + \beta_3 k_y^2) + k_x(\alpha_3 + \beta_4 k_y^2)}\right), \arctan\left(\frac{-3\beta_1 k_x k_y^2}{k_y(\alpha_2 + \beta_3 k_y^2) + k_x(\alpha_3 + \beta_4 k_y^2)}\right) \right)$$ and

$\left( \arctan\left(\frac{\gamma_2 k_x}{\gamma_1 k_y}\right), \arctan\left(\frac{\gamma_3 k_x}{\gamma_1 k_y}\right) \right)$, respectively. Then, a small deviation is expected to occur somewhere along the $k_y$ path away from both the Γ and Y points, *i.e.*, around the midway point of the $k_y$ path, because large $k_y$ minimizes the spin-deviation angles $(\theta_y, \theta_z)$. This conclusion is also consistent with the results derived by solving Eq. (2) (see Methods). Last, if the SOC parameters in the directions of $\sigma_y$ and $\sigma_z$ are also small, the angles $(\theta_y, \theta_z)$ would further decrease. If the third-order terms are ignored in Eq. (1), we can obtain spin-deviation angles that reduce to $(\arctan\left(\frac{\alpha_1 k_x}{\alpha_2 k_y + \alpha_3 k_x}\right), 0)$, which is similar to the spin deviation present in the $C_{2v}$ Hamiltonian. The PST phenomena were reported previously for materials with $C_{2v}$ and $C_s$ point groups. Here, the $k_y$ path having the mirror symmetry in the $C_{3v}$ point group resembles the points having $C_{2v}$ symmetry and point having $C_s$ symmetry. This simplification can be more effective when the *k* point is around the midway point of the $k_y$ path, where the spin deviation angles $(\theta_y, \theta_z)$ are also minimal.

**Validation of the $k \cdot p$ model**

We now apply these guidelines to materials with $C_{3v}$ point group symmetry and compute the spin textures using density functional theory (DFT) calculations. We first choose RbNbO$_3$ with *R*3*m* symmetry and LiTaO$_3$ with *R*3*c* symmetry, whose primitive structures in real and reciprocal space shown in **Figures 1a, 1b, 2a** and **2b**. Both compounds are insulators and have conduction band minimum (CBM) at Γ (**Figures 1c** and **2c**). The conduction band edges of both compounds indicate the PSTs occur along the $k_x$ direction (**Figures 1d** and **2d**), because there is a mirror symmetry in the $k_y$ path. The length of the arrows for the spin textures in **Figures 1d** and **2d** indicate that the spin amplitude along the $k_x$ direction in RbNbO$_3$ is larger than that in LiTaO$_3$. This result is consistent with our $\boldsymbol{k} \cdot \boldsymbol{p}$ model, which finds that the spins along the $k_y$ direction on the two Ta atoms with opposite directions are allowed and reduce the spins magnitude along the $k_x$ direction (*i.e.*, PST direction), when closer to the Γ point. Another conclusion from our $\boldsymbol{k} \cdot \boldsymbol{p}$ model analysis is that the PST can also occur at the midway point of the $k_y$ path from Γ to Y as shown in **Figures 1d** and **2d**. The PST areas with deviation angles (5°, 5°) for the two materials are enclosed by the orange lines. If we want to access the PST and its associated helix experimentally, one additional requirement must be satisfied. The PST region should be located around the CBM to enable its access through n-type doping. Although we find the PST in both RbNbO$_3$ and LiTaO$_3$ materials, this requirement is not satisfied. The PSTs will be difficult to access because the PST region is at a much higher energy than the CBM.

**Identification of PSTs in Tl$_3$SbS$_3$ and LiNbO$_3$**

We now apply our group theory analysis and $\boldsymbol{k} \cdot \boldsymbol{p}$ model to Tl$_3$SbS$_3$ with a polar $R3m$ ground-state structure [42] (**Figures 3a** and **3b**). Our computed electric polarization is approximately 23 $\mu$C cm$^{-2}$. The band structure of Tl$_3$SbS$_3$ reveals it has a DFT indirect band gap of 0.88 eV in bulk with the CBM along the A′ − Z path (**Figure 3c**). We find A′ − Z: (1/6,1/6,1/6) →(1/2,1/2, -1/2) at a $k_z = \frac{1}{2}$ (fractional coordinate) plane (**Figure 3c**), which exhibits a mirror symmetry perpendicular to *x*, can be transformed to a $[k, -k, 1/2]$ path in a hexagonal system. Following the group analysis above, the unidirectional spin direction should then be along the *x* direction since the mirror symmetry is perpendicular to *x*. Indeed, **Figure 3d** shows there is a PST with a unidirectional spin direction along the *x* direction for bands dispersing along the $k_y$ direction. Since the PST region is at a $k_z = \frac{1}{2}$ plane, the spin deviation is unaffected by $k_z$. Because space group *R*3*m* has 3-fold rotational symmetry, we expect three directions in the Brillouin zone exhibit a PST, which are symmetry-related to each other, and provide persistent spin helices (PSHs).

To determine the spin lifetime for the PSHs in Tl$_3$SbS$_3$, we adopt a Hamiltonian with $C_s$ symmetry, because the PST region is located around the middle of the $k_y$ path with a mirror operation. The SOC strength in the PST region along $k_y$ are $\gamma_1$=0.69 eV Å and those along $k_x$ are $\gamma_2$=0.06 eV Å and $\gamma_3$=0.71 eV Å. **Figure 3d** shows that the PST persists and spans an area from the CBM to 45 meV above it. Simultaneously, the spin deviation away from the *x* direction is also less than (5°, 5°). With the SOC parameters and a Fermi wavelength $k_F = 0.025$ Å$^{-1}$, we compute the spin lifetime $\tau_s \cong 1$ ps for the PSH mode and the characteristic spin lifetime is $\frac{\tau_s}{T_{PSH}}$ =8.8, where the spin procession period time $T_{PSH} = \frac{\pi\hbar}{\gamma_1 k_F}$. The spin lifetime in Tl$_3$SbS$_3$ is much shorter compared to canonical materials such as GaAs/AlGaAs (~200) [43].

To further enhance the quality of the PST in Tl$_3$SbS$_3$, we need to eliminate the influence of the SOC terms along $k_x$. Often when investigating Rashba, Dresselhaus, and persistent spin textures in quantum-well materials, **k** · **p** models in two-dimensions are invoked and the average momentum along the *out-of-plane* direction of a confined electron gas or thin film /heterostructure is set to zero within a mean-field approximation. The consequence is that odd order SOC terms in the Hamiltonian can be neglected. In a similar way, we build a thin film with $k_x$ along the *out-of-plane* direction and a mirror symmetry in the *in-plane* direction. Since $k_y$ and $k_z$ then become the *in-plane* reciprocal lattice vectors for the thin film, we explore the SOC terms along both $k_y$ and $k_z$ directions. Based on $C_s$ symmetry, we have $\gamma_1' k_z \sigma_x$ in $\mathcal{H}_Y$, which also supports the uniform spin direction along the $k_x$ direction same as that led by the SOC effects along the $k_y$ direction [Eq. (2)]. Therefore, a thin film on (011) TbAlO$_3$ substrate with its *out-of-plane* (*in-plane*) direction aligned along $k_x$ ($k_y$ and $k_z$) of the bulk structure (see **Figures 4a and 4b**) will support a uniform spin direction along the *out-of-plane* direction with zero spin deviation. This behavior is demonstrated by our direct DFT calculations (**Figure 4d**) with strains along the *a* and *b* directions of 1.8% and -0.2%, respectively. We find a perfect PST spanning a large portion of the Brillouin zone and a high energy range above the CBM.

If we further redefine the *in-plane* direction ($k_x'$) of the thin film to be along $\gamma_1 k_y + \gamma_1' k_z$ of the bulk state, which is a direction about 34° away from $k_z$ direction. Then, the CBM occurs along this direction in the thin film as seen in the $\Gamma - L$ path of the band structure (**Figure 4c**). The SOC parameter is 1.71 eV Å around the CBM along the $\Gamma - L$ path, which is comparable to other predicted PST materials with strong SOC coefficients, such as that of 1.9 eV Å in BiInO$_3$. We emphasize that the Tl$_3$SbS$_3$ film is the best three-dimensional material to the date showing a perfect PST without any spin deviation. Our finding stems from careful analyses of trigonal polar groups and reducing the $C_{3v}$ crystal class to $C_s$ by engineering an epitaxial thin film using demonstrated approaches [39] applied to realize monoclinic BiFeO$_3$. **Figure 4b** shows that a Tl$_3$SbS$_3$ film has *Cm* symmetry and the computed polarization for this phase is about 23 $\mu$C cm$^{-2}$ along the *b* direction. Because there is no spin deviation, the spin lifetime will ideally diverge toward infinity at low temperatures until another spin-scattering mechanism becomes operative. Alternative to the thin film geometry [44], this optimal PST may also be observed in bulk Tl$_3$SbS$_3$ along the $k_x$ direction via laser-induced formation of surface nanolayers [45].

A PSH state with the spin-spiral plane perpendicular to the unidirectional effective field has long been pursued [5]. By solving a microscopic spin-diffusion equation in quantum-well structures, previous studies found an enhanced spin lifetime $\tau_s$ occurs at a "magic" wavevector of $2q$ where $q$ corresponds to the shifting vector induced by the Kramers degeneracy from time reversal symmetry on the Fermi surface [1,5]. This well-known PSH state at $2q$ occurs as a consequence of the $C_{2v}$ or $C_s$ symmetry of the SOC Hamiltonian—the only two previously known symmetries of the SOC Hamiltonian that can produce the PSH state with a long spin lifetime [5,20]. This PSH mode is due to a PST spin texture that occurs in a $\boldsymbol{k}$ path starting from a high symmetry $k$ point having $C_{2v}$ or $C_s$ symmetry, resulting in a uniform spin direction that is in a direction perpendicular to the $\boldsymbol{k}$ path [31]. However, there exists a $\boldsymbol{k}$ path under $C_s$ symmetry, which may induce a spin texture: $k_x(\gamma_2\sigma_y + \gamma_3\sigma_z)$ such that $k_x$ does not contain a mirror plane. In this new defined spin texture, the spin texture in the plane perpendicular to $k_x$ depends on the ratio of $\gamma_2/\gamma_3$. This special situation can occur because of the SOC strengths along $k_y$ are smaller than those along the $k_x$ direction. Therefore, we call the PST along a $k$ path without mirror symmetry as a Type-II or accidental PST to contrast it with Type-I PSTs that are enforced along a $k$ path with a mirror symmetry.

Next, we will elucidate how a Type-II PST occurs in the important optoelectronic material LiNbO$_3$ [46] (**Figure 5a**), for which we calculate an electric polarization of 68 $\mu$C cm$^{-2}$. Our computed polarization is slightly underestimated compared with the experimentally observed polarization [47] (77 $\mu$C cm$^{-2}$). **Figure 5c** shows that the CBM is not along a high symmetry $k_y$ path having the mirror symmetry but is at different $k$ point comparted to LiTaO$_3$ where the CBM is in the $k_y$ path. The location of the CBM in the LiNbO$_3$ is consistent with the previous calculations [48]. Since we are interested in the $k_x - k_y$ plane for searching for possible PST regions, we construct a thin film LiNbO$_3$ geometry with *Cc* symmetry ($a \approx b$=3.68Å) (see **Figure 5d**), which should be accessible in experiment using demonstrated growth approaches [39] to realize *Cc* BiFeO$_3$. The computed polarization for the film is 71 $\mu C$ cm$^{-2}$ along a direction 37° from the *c* direction in the (110) plane. We find that the CBM is located at a point in the $k_x$ path along [110] direction in the thin film. We further interpolate the band structure in a $k_x - k_y$ plane at $k_z = 0$ by carrying out Wannier90 + SOC calculations (**Supplementary Figure 1**), which further supports our identification of the CBM the minimum energy along the $k_x$ path. Surprisingly, although the $k_x$ path only possesses the identify operation, there is still a region showing uniform spin direction determined by $\gamma_2 \sigma_y + \gamma_3 \sigma_z$. This Type II PST region also has a sizable area as shown in **Figure 5f**, enclosed by the (orange) boundary line demarcating the spin deviation from a direction determined by $\gamma_2 \sigma_y + \gamma_3 \sigma_z$ (*i.e.*, ~arctan($\gamma_3/\gamma_2$)=33° from the $k_y$ direction) is also less than 10°. The SOC strength in the PST region along $k_y$ is $\gamma_1$=0.14 eV Å and along $k_x$ it is $\gamma_2$=0.51 eV Å and $\gamma_3$=0.33 eV Å in the *Cc* structure. With these SOC parameters and a Fermi wavelength of 0.04 Å$^{-1}$, we compute the spin lifetime for the PSH mode in LiNbO$_3$ to be 11 ps and $\frac{\tau_s}{T_{PSH}}$ =127. Therefore, a possibly better PST in the polar structure with *C*$_{3v}$ symmetry can be obtained by reducing its symmetry to *C*$_s$ in a thin film, while there is no PST accessible in bulk *R*3*c* structure.

**Conclusion**

We proposed a strategy to identify and design optimal PSTs in bulk crystals based on polar space groups showing an odd number of mirror operations. We showed that compounds with point groups *C*$_{3v}$ may also exhibit PSTs, which expands the phenomenon to more readily n-type dopable chalcogenide compounds and perovskite oxides with *R*3*c* and *R*3*m* symmetries. We also found that using strain to reduce the crystalline symmetry is a useful strategy for improving the performance of bulk PSTs and unlocking hidden Type-II (or accidental) PSTs in thin films of many previously identified Rashba compounds [23,42]. This approach brings PST properties to more complex crystal structures and chemistries with strong Rashba coefficients and/or topological insulator and Weyl semimetal phases. Noncentrosymmetric compounds with high symmetry wavevectors in reciprocal space exhibiting $C_s$ point symmetry are an important initial phase space to search for PSTs in known materials. Last, our study demonstrates a new type of PST that does not require symmetry protection, which will bring further opportunities to find high quality PSHs in future spin-orbitronic devices.

## Acknowledgements

X.-Z.L. and J.M.R. were supported by the National Science Foundation (NSF) under DMR-2104397. DFT calculations were performed on the CARBON cluster at the Center for Nanoscale Materials at Argonne National Laboratory and the Extreme Science and Engineering Discovery Environment (XSEDE), which is supported by NSF (ACI-1548562). Use of the Center for Nanoscale Materials, an Office of Science user facility, was supported by the U.S. Department of Energy, Office of Science, Office of Basic Energy Sciences, under Contract No. DE-AC02-06CH11357.

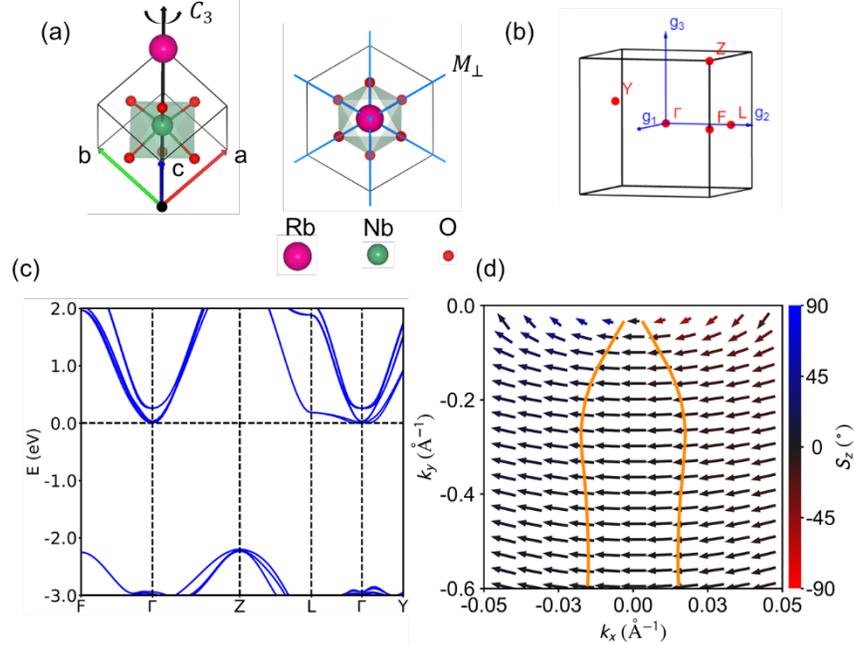

**Figure 1. a** The trigonal cell of RbNbO$_3$ with $R3m$ symmetry along the 3-fold axis. The mirror symmetries are also shown. **b** The Brillouin zone of the trigonal cell reciprocal lattice vectors (g$_1$, g$_2$, g$_3$). The high symmetry $k$ points are shown in red with values specified in Supplementary Table 11. **c** The band structure of the trigonal cell. The energy is given with respect to the CBM. **d** The spin textures in the $k_z = 0$ (fractional coordinate) plane of the lowest conduction band around the $\Gamma - Y$ path. The arrows indicate the spin direction. The color scale represents the degree of spin deviation out of the $xy$ plane (i.e., along the $k_z$ direction). The area within the orange lines indicates spin deviations less than (5°, 5°).

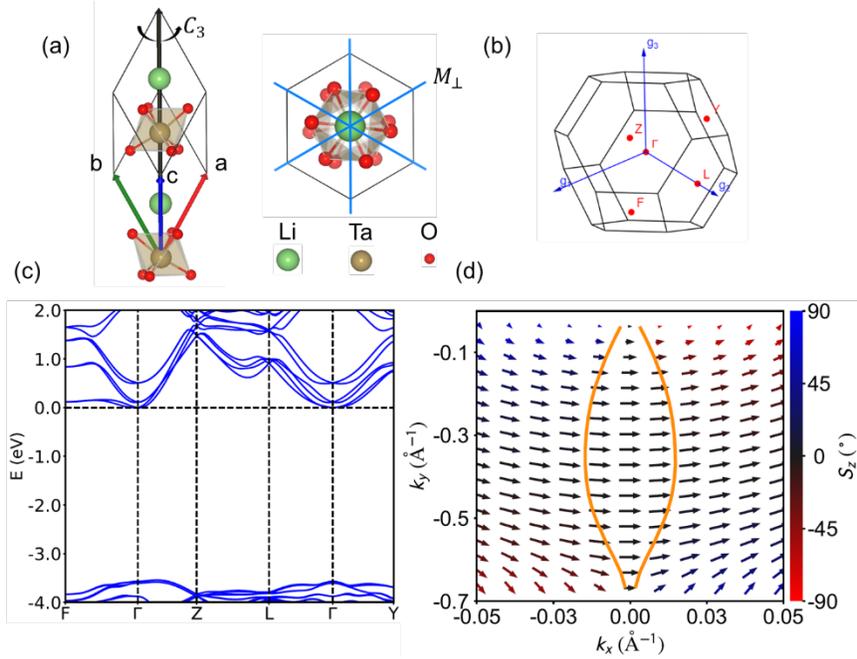

**Figure 2. a** The trigonal cell of LiTaO₃ with *R*3c symmetry along the 3-fold axis. The mirror symmetries are also shown. **b** The Brillouin zone of the trigonal cell with reciprocal lattice vectors ($g_1$, $g_2$, $g_3$). The high symmetry *k* points are shown in red with values specified in Supplementary Table 11. **c** The band structure of the trigonal cell. The energy is given with respect to the CBM. **d** The spin textures in the $k_z = 0$ (fractional coordinate) plane of the lowest conduction band around the Γ − Y path. The arrows indicate the spin direction. The color scale represents the degree of spin deviation out of the *xy* plane (i.e., along the $k_z$ direction). The area within the orange lines indicates spin deviations less than (5°, 5°).

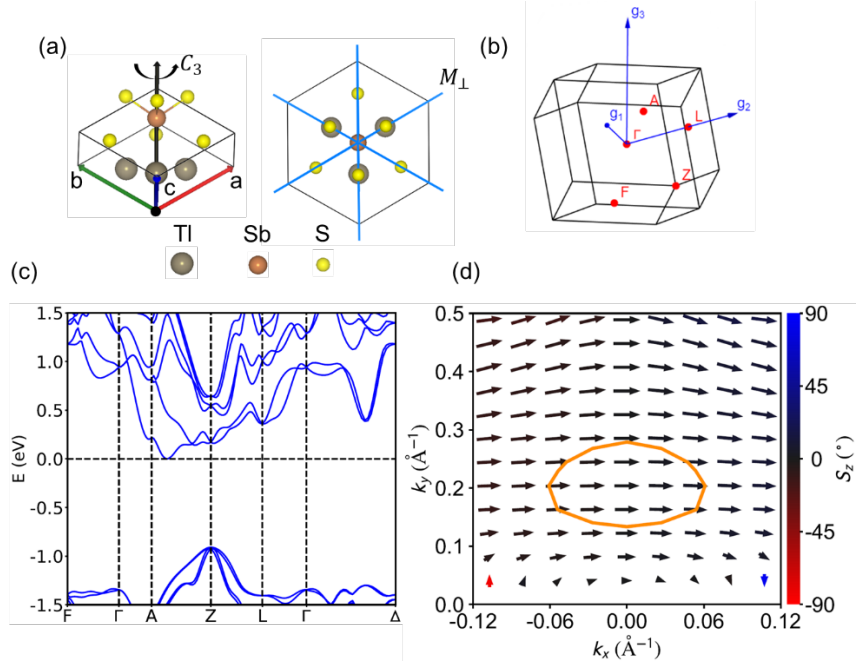

**Figure 3. a** The trigonal cell of Tl$_3$SbS$_3$ with $R3m$ symmetry along the 3-fold axis. The mirror symmetries are also shown. **b** The Brillouin zone of the trigonal cell with reciprocal lattice vectors ($g_1$, $g_2$, $g_3$). The high symmetry $k$ points are shown in red and specified in Supplementary Table 12. **c** The band structure of the trigonal cell. The energy is given with respect to the CBM. **d** The spin textures in the $k_z = 1/2$ (fractional coordinate) plane of the lowest conduction band around the $A - Z$ path. The arrows indicate the spin direction. The color scale represents the degree of spin deviation out of the $xy$ plane (i.e., along the $k_z$ direction). The area within the orange lines indicates spin deviations less than (5°, 5°).

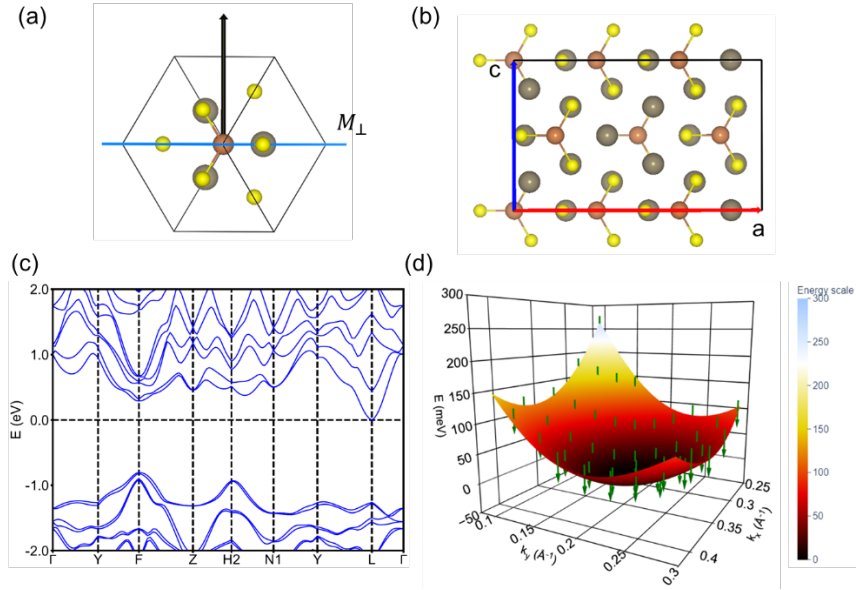

**Figure 4. a** Orientation for building thin films from $Tl_3SbS_3$ with a trigonal cell and $R3m$ symmetry. The black arrow indicates the film direction. **b** The structure of $Tl_3SbS_3$ film with $Cm$ symmetry. The Brillouin zone of the film is the same as that of the trigonal cell. The values of the high symmetry $k$ points are specified in Supplementary Table 13. **c** The band structure of the film. The energy is given with respect to the CBM. **d** The energy surface of the lowest conduction band spanning a $k_x - k_y$ plane around the A − Z path. The arrows indicate the spin direction. The color scale represents the band energies. All the spins are oriented along the $k_z$ direction.

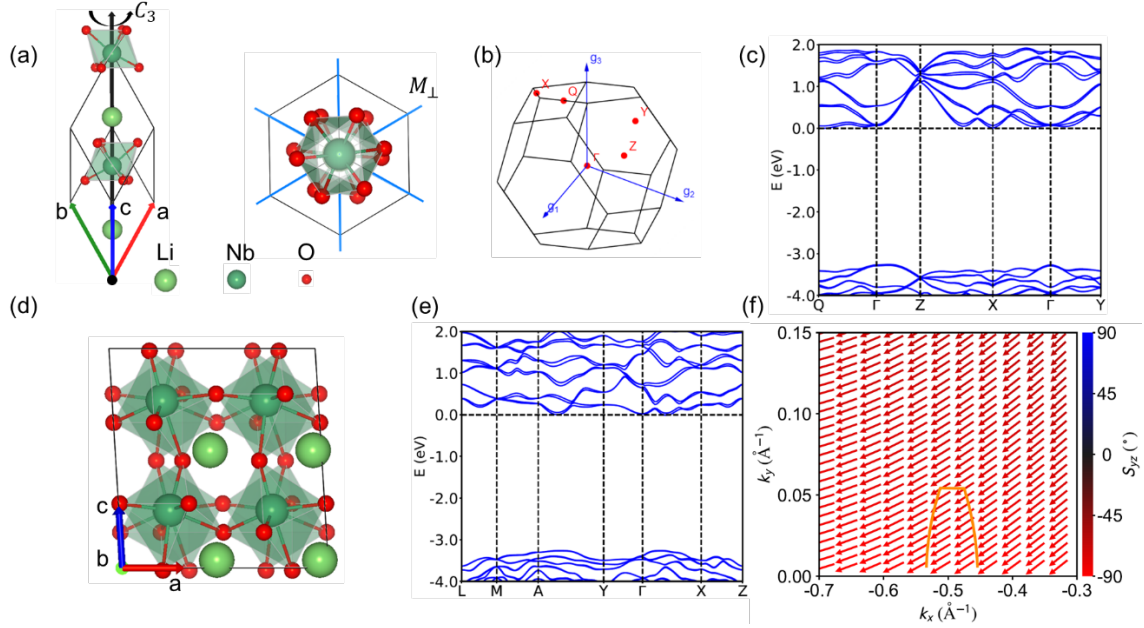

**Figure 5. a** The trigonal cell of LiNbO$_3$ with $R3c$ symmetry along the 3-fold axis. The mirror symmetries are also shown. **b** The Brillouin zone of the trigonal cell with reciprocal lattice vectors ($g_1$, $g_2$, $g_3$). The high symmetry $k$ points are shown in red with values specified in Supplementary Table 14. **c** The band structure of the trigonal cell. The energy is given with respect to the CBM. **d** The structure of a LiNbO$_3$ film with $Cc$ symmetry. The Brillouin zone of the film is same as that of the trigonal cell. The values of the high symmetry $k$ points are shown in Supplementary Table 15. **e** The band structure of the film. **f** The spin textures in the $k_z = 0$ (fractional coordinate) plane of the lowest conduction band around the $\Gamma - Y$ path. The arrow indicates the spin direction. The color scale represents the degree of spin deviation out of the $xy$ plane (i.e., along the $k_z$ direction). The area within the orange lines indicates spin deviations less than (5°, 5°). The arrow indicates the spin direction (in the $yz$ plane) projected into the $k_z = 0$ plane, whose components along $k_x$ and $k_y$ in the plot are along the Cartesian $y$ and $z$ directions, respectively. The color scale represents the degree of spin deviation out of the $yz$ plane (i.e., along the $k_x$ direction), that is, 90° indicates the spin is in the $yz$ plane. The area within the orange lines indicates spin deviation towards the $k_x$ direction is less than 10°.